# Inferring SQL Queries Using Program Synthesis

Alvin Cheung    Armando Solar-Lezama    Samuel Madden

MIT CSAIL

{akcheung,asolar,madden}@csail.mit.edu

## Abstract

Developing high-performance applications that interact with databases is a difficult task, as developers need to understand both the details of the language in which their applications are written in, and also the intricacies of the relational model. One popular solution to this problem is the use of object-relational mapping (ORM) libraries that provide transparent access to the database using the same language that the application is written in. Unfortunately, using such frameworks can easily lead to applications with poor performance because developers often end up implementing relational operations in application code, and doing so usually does not take advantage of the optimized implementations of relational operations, efficient query plans, or push down of predicates that database systems provide.

In this paper we present QBS, an algorithm that automatically identifies fragments of application logic that can be pushed into SQL queries. The QBS algorithm works by automatically synthesizing invariants and postconditions for the original code fragment. The postconditions and invariants are expressed using a theory of ordered relations that allows us to reason precisely about the contents and order of the records produced even by complex code fragments that compute joins and aggregates. The theory is close in expressiveness to SQL, so the synthesized postconditions can be readily translated to SQL queries. Using 40 code fragments extracted from over 120k lines of open-source code written using the Java Hibernate ORM, we demonstrate that our approach can convert a variety of imperative constructs into relational specifications.

## 1. Introduction

Almost all modern, interactive web applications store their persistent state in a relational database. Developing such database-backed applications is complicated because databases are frequently accessed using SQL, while most application programs are written in a high-level, often imperative, language such as Java or Python. These languages usually provide APIs to connect to databases and allow developers to retrieve and store persistent data by embedding SQL queries within their application code. Unfortunately, this approach has a number of issues. First, because the developer expresses her queries in SQL, she is forced to learn a different programming language (and often a different programming paradigm) in order to work with the database. Second, developers must flatten objects into relational tables, and marshall and unmarshall objects into SQL queries and out of query results. Third, this approach breaks the abstraction barrier between the database and application programs, since changes to the object hierarchy might involve changes to the embedded SQL queries or schemas.

A popular approach to address this so-called "impedance mismatch" between imperative and declarative interfaces is to use systems that allow developers to access database and non-database objects with the same general purpose language that the application is written in. This notion is sometimes referred to as *transparent persistence* and implemented in object-oriented databases [23], and object-relational mappings (ORM) libraries such as Hibernate [3] and JPA [2]. To use such libraries, the developer first designs objects that are intended to be persistently stored in a database using special annotations or external mapping files. The library provides a tool that takes in the user's design and creates the appropriate tables in the database, and abstracts the SQL queries that retrieve and update persistent objects in the database as API calls that the application can invoke.

Thus, the developer neither needs to learn a separate programming model nor understand how the database processes the queries that are issued by the application. Such libraries have become extremely popular; for example, as of July, 2012, on the job board dice.com 15% of the 17,000 Java developer jobs are for programmers with Hibernate experience.

Although popular, such frameworks often lead to inefficient use of the database. This is because developers often end up implementing many relational operations—e.g., filters, joins, and aggregates—in application code using loops over a set of records retrieved earlier from the database. This is due to programmers' lack of knowledge about relational operations, and/or because the interfaces for accessing complex relational operations are difficult to use in these languages. Running such operations in the application is very inefficient, because it does not allow databases to effectively optimize queries by using specialized operators or indices, and can result in shipping a large amount of unnecessary data to the application. For example, database systems often have a number of join implementations and can select the best one depending on the size and ordering of the input tables. In contrast, programmers using ORMs often implement application-level joins with nested loops, which can be several orders of magnitude slower than a database's optimized join implementations as our measurements show. In fact, anecdotal evidence suggests that when application performance becomes critical, developers using ORM layers often by-pass the ORM and access the database directly by issuing queries encoded with strings via its native SQL or SQL-like interface—often provided as a "backdoor" by the ORM libraries.

In this paper, we develop a new compiler analysis algorithm, QBS (Query By Synthesis), that allows developers using transparent persistence frameworks to continue writing applications as usual, implementing logic that could be pushed into the database in the application. Then, the QBS algorithm analyzes the code and automatically converts portions of it into SQL expressions that run in the database. Our algorithm retains the best of both worlds: developers do not need to understand the intricacies of SQL but can still develop efficient applications. This problem statement is not new: Cook *et al.* [32, 33] first identified this as the *query extraction problem* in a seminal 2007 POPL paper. Our paper significantly expands on this prior work by extending the subset of source programs that can be analyzed, and the expressiveness of the SQL queries that are generated. Specifically, to the best of our knowledge, our work is the first that is able to identify joins and aggregates in general purpose application logic and convert them to SQL queries. Our analysis ensures that the generated queries are *precise* in that both the contents and the order of records in the generated queries are the same as that produced by the original code.

Our basic approach is to infer an SQL expression that is semantically equivalent to a given block of imperative code, and to replace the imperative code fragment with the inferred SQL expression. The key idea is to recognize that this problem is equivalent to coming up with a postcondition for the imperative code block that can be translated into SQL. Our approach follows the classical results from axiomatic semantics [13] to compute verification conditions for the imperative code fragment symbolically in terms of the unknown postconditions and loop invariants. The resulting formula is fed to a constraint-based synthesis engine to solve for the unknown postconditions and invariants. Our synthesis approach generates postconditions and invariants using a carefully designed predicate language that can be readily compiled to SQL. If a postcondition can be found that makes the verification condition valid, then the corresponding imperative code fragment is converted into its relational specification in SQL.

In summary, this paper makes the following contributions:

1. We designed a predicate language that is based on a new theory of ordered relations that we developed. The language is designed to bridge the gap between the imperative and relational worlds. Ordered relations are similar to standard relations, except that each record in the ordered relation is indexed. The theory is expressive enough to model various relational operations, and at the same time allows us to infer relational specifications for a variety of imperative program constructs. Using the theory, we are able to prove that the transformations are sound and fully precise.

2. Based on the theory, we develop the QBS algorithm based on automatic program verification. The algorithm takes in a code fragment written in an imperative language that uses an ORM library to retrieve records from database, automatically generates a relational specification from it, and converts it into an SQL query. Our technique solves the postcondition and loop invariant inference problem by analyzing the input program and inferring a grammar that describes the likely structure of the unknown predicates. This structural hypothesis is fed to a constraint-based synthesizer [28] that finds predicates that are likely to be correct based on bounded analysis. Our technique then uses Z3 [5] to prove that the verification condition is valid with respect to our theory of ordered relations.

3. We implemented a prototype of our algorithm and evaluated it on Java programs that use the Hibernate ORM library for data persistence operations. We demonstrate the feasibility of our algorithm by inferring relational specifications from 40 different real-world code examples. The experiments show that our algorithm is able to infer significantly more classes of relational specifications compared to what was possible with prior work.

More generally, our system demonstrates the use of constraint-based synthesis technology to provide a general solution to what had been previously considered a difficult compiler analysis problem, and the theory that we developed also provides a new way to reason about relational operations in an imperative context.

## 2. Overview

In this section we describe the overall operations of the QBS algorithm, starting with an example of it running on an excerpt from an open source project management application [4] written using the Hibernate framework. Figure 1 shows the code fragment for the example.

```
1  List<User> getRoleUser () {
2    List<User> listUsers = new ArrayList<User>();
3    List<User> users = this.userDao.getUsers();
4    List<Role> roles = this.roleDao.getRoles();
5    for (int i = 0; i < users.size(); i++) {
6      for (int j = 0; j < roles.size(); j++) {
7        if (users.get(i).roleId().equals(roles.get(j).roleId())) {
8          User userok = users.get(i);
9          listUsers.add(userok);
10       }
11     }
12   }
13   return listUsers;
14 }
```

**Figure 1.** Sample code that implements join operation in application code, abridged from actual source for clarity

**Postcondition**

$listUsers = \pi(\text{sort}(\sigma(\bowtie (users, roles, \text{True}), f_\sigma), l_{sort}), f_\pi)$
where
$f_\sigma := \text{get}(users, i).roleId = \text{get}(roles, j).roleId$
$f_\pi$ projects all the fields from the *User* class
$l_{sort} = [users, roles]$

**Translated code**
```
1  List<User> getRoleUser () {
2    List<User> listUsers = db.executeQuery(
3      "SELECT u
4       FROM users u, roles r
5       WHERE u.roleId == r.roleId
6       ORDER BY u.roleId, r.roleId");
7    return listUsers; }
```

**Figure 2.** Postcondition as inferred from Fig. 1 and code after query transformation

In this example, the method `getRoleUser` is used to return the list of users whose role is currently stored in the database. It first populates the list `users` on line 3 using a call to the Hibernate library to retrieve all `User` objects that are currently stored in the database, and similarly for the `roles` list. The code then iterates through two loops to compare each user against each role, and adds those users that pass the criteria to `listUsers` on line 9. Finally, this list is returned as the result of the function.

While the example implements the desired functionality and is easy to read, it performs poorly. First, when the method is executed, the ORM library needs to fetch all the records from the corresponding tables that store `User` and `Role` objects using a SQL query, marshall each record into a Java object, and return the results to the user application as lists. However, since only a subset of `users` (and none of `roles`) are actually returned, resources are wasted in fetching and marshalling the extra database records that are not needed. This problem is exacerbated when the sizes of the Users and Roles relations are large when compared to the size of the returned results. Meanwhile, notice that `getRoleUser` essentially implements a relational join and projection operation in application code, but without utilizing indices or efficient join algorithms the database system may have access to.

The goal of the QBS algorithm is to compile the application to the code shown in the bottom of Fig. 2, where the nested loop is converted to a SQL query that implements the same functionality that is assigned to the variable `listUsers` as in the original code. The algorithm works by first identifying the program variable that the results from the inferred query should be assigned to (in the case of the running example it is `listUsers`) — we refer this as the "postcondition variable." In the current implementation we mark either the return value of the input code fragment as the postcondition variable, or ask the user to indicate the result variable that our algorithm should be applied to using a special annotation (inferring the postcondition variable automatically is a topic for future work). The body of the function prior to the location where the postcondition variable is marked is then considered for SQL transformation. We currently assume that all database records that are used in the code fragment to be analyzed are fetched by the code fragment. For example, we assume that they are not passed in as function parameters.

Once we have identified the block of code we are going to infer an equivalent SQL query for, we need to infer an expression for the postcondition variable that is implied by the code fragment and is translatable to SQL. In the case of the running example, the translated result is shown in line 3 in Fig. 2. We describe these steps in more detail in the rest of this section.

**Conversion to kernel language representation.** Given the input code fragment, the first step in the QBS algorithm is to convert the code fragment into a simplified kernel language. The description of the kernel language and the process of translation from Java source into this language are discussed in Sec. 4.

**Generation of verification conditions.** As the next step, we compute the verification conditions of the program expressed in the kernel language, without assuming any preconditions when the code fragment is entered. The verification conditions are written using the predicate language derived from the theory of ordered relations to be discussed in Sec. 3, and the procedure used to compute verification conditions is a fairly standard one [13, 16]. The interesting point here is that at this point we do not know what the postcondition and loop invariants (if any) are. Thus, during this initial verification process we symbolically represent the verification conditions in terms of unknown invariants and postcondition. The process of computing verification conditions is discussed in more detail in Sec. 5.

**Synthesis of invariants and postcondition.** The definitions of the postcondition and invariants need to be filled in and validated before translation can proceed. We do this using a synthesis-based approach that is similar to [29], where we use a synthesizer to come up with a postcondition that is implied by the computed verification conditions. During synthesis, the synthesizer performs bounded verification on the postcondition candidates and only returns those that can be successfully verified. To prevent the synthesizer from generating trivial postconditions (such as True), we limit the synthesizer to only generate postconditions that can be translated to SQL, as defined by our theory of ordered relations. If there are loops in the input code fragment, we ask the synthesizer to generate loop invariants as well. We provide a template, written in the predicate language, that describes the space of possible postcondition and invariants to the synthesizer. For instance, from the example shown in Fig. 1, our algorithm infers the postcondition shown at the top of Fig. 2, where sort, $\pi$, $\sigma$, and $\bowtie$ refers to sorting, relational projection, selection, and join, respectively. The process of automatic template generation from the input code fragment and synthesis of the postcondition from the template are discussed in Sec. 6.

Determining loop invariants is undecidable for arbitrary programs in general [7]. In our case, however, we do not need to determine the strongest invariants or postconditions: we are only interested in finding postconditions that allow us transform the input program into an SQL expression. In the case of the running example, we are only interested in finding a postcondition of the form listUsers = Query(q), where q is a SQL query to be issued to the database. Similarly, we only need to discover loop invariants that are strong enough to prove the postcondition of interest. This insight greatly reduces the space of solutions that we need to consider and also shortens the time needed to find a solution, as our experiments show in Sec. 8.

**Validation and SQL conversion.** Once the synthesizer finds a solution to the postconditions and invariants in the verification conditions, the candidates are sent to a theorem prover for verification, since the synthesizer used in our algorithm is only able to perform bounded verification. If the prover returns a positive answer, the input program is translated to SQL, as shown in the bottom of Fig. 2. The predicate language defines rules to translate any expressions in the language into valid SQL. The details of validation is discussed in Sec. 7 while the rules for SQL conversion are introduced in Sec. 3.4.

We now turn to a discussion of the theory of ordered relations, which form the basis of our approach. Once we have presented this theory, we describe the above steps in more detail, and evaluate our approach.

## 3. Theory of Finite Ordered Relations

The theory used to reason about the postconditions and invariants in the imperative code is central to our ability to infer relational specifications from imperative programs. The theory must allow *precise* analysis of the imperative code to enable reasoning about both the set of records produced and the ordering of records. This is important since many source languages (such as Java) expose an ordered list interface to manipulate records retrieved from the database, and subsequent operations in the client code might rely on the order in which the records are returned.

There are other ways to model relational operations in imperative code (see Sec. 9 for details). However, they are either limited in expressiveness for reasoning about order-preserving relational operations, or cannot be easily translated to SQL. In this section, we describe a new theory of finite ordered relations that the QBS algorithm uses along with the translation of the operators in the theory into SQL queries.

### 3.1 Basics

The theory of finite ordered relations is based on the theory of lists with standard list operations. The theory operates on three types of values: scalars, records, and ordered relations of finite length. Records are collections of named fields, and an ordered relation is a finite list of records. Each record in the relation is labeled with an integer index that can be used to fetch the record. The theory defines a number of operators that can be used to manipulate ordered relations. Figure 3 presents the abstract syntax of the theory and shows how to combine operators to form expressions. The theory assumes that all the operators are applied to values of the right type.

The operators introduced in the language mirror the operators of standard relational algebra. The operators fall into two categories: those that return an ordered relation, and those that return a single record. The first group includes projection of fields from records ($\pi$), selection of records ($\sigma$), joining of ordered relations ($\bowtie$), sort, unique, and top. Those that return a single record include get, size, contains, max, min, and sum. The expression Query represents the original query; it retrieves records from the database and returns them as an ordered relation.

### 3.2 Axioms

The semantics of the operators in the theory are defined recursively by the axioms in Fig. 4. The operator size($r$) returns the number of records in relation $r$, and get($r$, $i$) returns the record that is stored

## Figure 4. Axioms in the theory of ordered relations

**size**
$$\text{size}([\,]) = 0 \qquad \text{size}(r) = 1 + \text{size}(\text{tail}(r))$$

**get**
$$\begin{aligned}
i = 0 &\rightarrow \text{getHelper}(h:t, i) = h \\
i > 0 &\rightarrow \text{getHelper}(h:t, i) = \text{getHelper}(t, i-1) \\
\text{get}(r, i) &= \text{getHelper}(r, \text{size}(r) - i - 1)
\end{aligned}$$

**append**
$$\begin{aligned}
\text{append}([\,], r) &= r & \text{append}(h:t, r) &= h : \text{append}(t, r) \\
\text{append}(r, e) &= r : e & \text{append}([\,], e) &= [e]
\end{aligned}$$

**top**
$$\text{top}([\,], i) = [\,] \qquad \text{top}(r, i) = \text{top}(\text{tail}(r), i-1) : \text{get}(r, i-1)$$

**join ($\bowtie$)**
$$\begin{aligned}
\bowtie (r, [\,], f) &= [\,] & \bowtie (h:t, r, f) &= \bowtie_l (h, r) : \bowtie (t, r, f) \\
\bowtie ([\,], r, f) &= [\,] & \bowtie (r, h:t, f) &= \bowtie_r (r, h) : \bowtie (r, t, f)
\end{aligned}$$

$$\begin{aligned}
f(h, e) = \text{True} &\rightarrow \bowtie_l (h:t, e, f) = (h, e) :\bowtie_l (t, e, f) \\
f(h, e) = \text{False} &\rightarrow \bowtie_l (h:t, e, f) = \bowtie_l (t, e, f)
\end{aligned}$$

$$\begin{aligned}
f(e, h) = \text{True} &\rightarrow \bowtie_r (e, h:t, f) = (e, h) :\bowtie_r (e, t, f) \\
f(e, h) = \text{False} &\rightarrow \bowtie_r (e, h:t, f) = \bowtie_r (e, t, f)
\end{aligned}$$

**projection ($\pi$)**
$$\pi([\,], f) = [\,] \qquad \pi(h:t, f) = f(h) : \pi(t, p)$$

**selection ($\sigma$)**
$$\sigma([\,], f) = [\,]$$
$$\begin{aligned}
f(h) = \text{True} &\rightarrow \sigma(h:t, f) = h : \sigma(t, f) \\
f(h) = \text{False} &\rightarrow \sigma(h:t, f) = \sigma(t, f)
\end{aligned}$$

**sum**
$$\text{sum}([\,]) = 0 \qquad \text{sum}(h:t) = h + \text{sum}(t)$$

**max**
$$\text{max}([\,]) = -\infty$$
$$\begin{aligned}
h > \text{max}(t) &\rightarrow \text{max}(h:t) = h \\
h \leq \text{max}(t) &\rightarrow \text{max}(h:t) = \text{max}(t)
\end{aligned}$$

**min**
$$\text{min}([\,]) = \infty$$
$$\begin{aligned}
h < \text{min}(t) &\rightarrow \text{min}(h:t) = h \\
h \geq \text{min}(t) &\rightarrow \text{min}(h:t) = \text{min}(t)
\end{aligned}$$

**contains**
$$\text{contains}(e, [\,]) = \text{False}$$
$$\begin{aligned}
e = h &\rightarrow \text{contains}(e, h:t) = \text{True} \\
e \neq h &\rightarrow \text{contains}(e, h:t) = \text{contains}(e, t)
\end{aligned}$$

---

$$\begin{aligned}
c \in \text{constant} &::= \text{True} \mid \text{False} \mid \text{number literal} \mid \text{string literal} \\
e \in \text{expression} &::= c \mid \text{program var} \mid \{f_i = e_i\} \mid e_1 \text{ op } e_2 \mid \neg e \\
&\mid [\,] \mid \text{Query}(\ldots) \mid \text{size}(e) \mid \text{get}(e_r, e_s) \\
&\mid \text{top}(e_r, e_s) \mid \pi(e, f_\pi) \mid \sigma(e, f_\sigma) \\
&\mid \bowtie (e_1, e_2, f_\bowtie) \mid \text{append}(e_1, e_2) \\
&\mid \text{sort}(e, [e.f_i]) \mid \text{unique}(e) \\
&\mid \text{sum}(e) \mid \text{max}(e) \mid \text{min}(e) \\
f_\pi(e) &::= \{e.f_i\} \\
f_\sigma(e_s) &::= \wedge\ e.f_i\ \text{op } c \mid e.f_i\ \text{op } e.f_j \mid \text{contains}(e_s, e) \\
f_\bowtie(e_1, e_2) &::= \wedge\ e_1.f_i\ \text{op } e_2.f_j \\
\text{op} \in \text{binary op} &::= \wedge \mid \vee \mid > \mid < \mid \geq \mid \leq \mid =
\end{aligned}$$

**Figure 3.** Abstract syntax for the predicate language based on the theory of ordered relations

---

at index $i$ of relation $r$. append($r_1, r_2$) inserts relation $r_2$ at the end of $r_1$ and returns a new ordered relation. top($r, i$) returns a relation containing the first $i - 1$ records in the relation $r$.

The axioms for $\pi$, $\sigma$ and $\bowtie$ are modeled after relational projection, selection, and join respectively, but they also define an order for the records in the output relation relative to those in the input relations. The $\pi$ operator performs projection of fields from an ordered relation. The operator passes each record in the relation to the function $f_\pi$, which selects a number of fields and creates a new record using those fields. Just like the equivalent operation in relational algebra, the same field can be replicated multiple times; $f_\pi$ simply renames the new fields as needed. The $\sigma$ operator uses a function $f_\sigma$ to filter records from the input relation. $f_\sigma$ is a boolean function defined as a conjunction of boolean predicates, where each predicate either compares the value of a record field and a constant, the values of two record fields, or checks if the record is contained in another relation using contains.

The $\bowtie$ operator implements the relational join operation between two ordered relations. The operator iterates over each record $h_1$ from the first relation and pairs that with each record $h_2$ from the second relation. The two records are passed to the boolean function $f_\bowtie$, and the pair is added to the resulting relation if $f_\bowtie$ returns True. $f_\bowtie$ is similar to $f_\sigma$, except that it is a conjunction of boolean predicates of the form $e_1.f_i$ op $e_2.f_j$, i.e., each predicate compares values of one of the fields from the first relation to another field from the other relation. The axioms that define the aggregate operators max, min, and sum assume that the input relation contains only one numeric field, namely the field to aggregate upon.

The definitions of unique and sort are standard; in the case of sort, the second argument is a list of fields to sort by. Our system does not actually reason about these two operations in terms of their definitions; instead it treats them as uninterpreted functions with certain algebraic properties outlined in the following section. Because of this, there are some formulas involving sort and unique that we cannot prove, but we have not found this to be significant in practice (see Sec. 8 for details).

### 3.3 Operator Equivalence

The axioms that define the basic operations imply a number of properties that will be useful when translating expressions in the predicate language into SQL.

**Theorem 1 (Operator Equivalence).** *The following equivalences hold, both in terms of the contents of the relations and also the ordering of the records in the relations:*

- $\sigma(\pi(r, f_\pi), f_\sigma) = \pi(\sigma(r, f_\sigma), f_\pi)$
- $\sigma(\sigma(r, f_{\sigma_1}), f_{\sigma_2}) = \sigma(r, f')$, where $f' = f_{\sigma_2} \wedge f_{\sigma_1}$
- $\text{top}(\sigma(r, f), e) = \sigma(\text{top}(r, e), f)$
- $\pi(\pi(r, f_{\pi_1}), f_{\pi_2}) = \pi(r, f')$, where $f'$ is the concatenation of all the fields in $f_{\pi_1}$ and $f_{\pi_2}$.
- $\text{top}(\pi(r, f), e) = \pi(\text{top}(r, e), f)$
- $\text{top}(\text{top}(r, e_1), e_2) = \text{top}(r, \max(e_1, e_2))$
- $\bowtie (r_1, r_2, f_\bowtie) = \sigma(\bowtie (r_1, r_2, \text{True}), f_\bowtie)$, i.e., joins can be converted into cross products with selections.
- $\bowtie (\text{sort}(r_1, l_1), \text{sort}(r_2, l_2), f) = \text{sort}(\bowtie (r_1, r_2, f), l_1 : l_2)$
- $\bowtie (\pi(r_1, f_{\pi_1}), \pi(r_2, f_{\pi_2}), f) = \pi(\bowtie (r_1, r_2, f), f')$, where $f'$ is the concatenation of all the fields in $f_{\pi_1}$ and $f_{\pi_2}$.

Except for the equivalences involving sort, the other ones can be proven easily from the axiomatic definitions.

### 3.4 Translating to SQL

The expressions defined in the predicate grammar can be converted into semantically equivalent SQL queries. In this section we prove that any expression that does not use append or unique can be compiled into an equivalent SQL query. We prove this in two steps; first, we define *translatable expressions* and show that any expression that does not use append or unique can be converted into a translatable expression using the equivalences in Thm. 1. Then we show how to produce SQL from translatable expressions.

**Definition 1** (**Translatable Expressions**). *The following grammar defines the set of expressions that can be directly translated into SQL:*

$$
\begin{aligned}
b \in \text{baseExp} &::= \text{Query}(...) \mid \text{top}(s, e) \mid \bowtie (b_1, b_2, \text{True}) \mid \text{agg}(t) \\
s \in \text{sortedExp} &::= \pi(\text{sort}(\sigma(b, f_\sigma), [e.f_i]), f_\pi) \\
t \in \text{translatableExp} &::= s \mid \text{top}(s, e)
\end{aligned}
$$

*where the term* agg *in the grammar denotes any of the aggregation operators (*min, max, sum, size*), and the other operators are the same as those presented in Fig. 3.*

**Theorem 2** (**Completeness of Translation Rules**). *All expressions in the predicate grammar in Fig. 3, except for those that contain any* append *or* unique *operators, can be converted into translatable expressions.*

*Proof.* We prove the theorem by defining an algorithm to perform the conversion. The algorithm is defined in terms of a function Trans defined in Fig. 5. The function takes in an expression whose sub-expressions are all translatable expressions and produces an equivalent translatable expression. By applying this function to the AST of an arbitrary expression that does not use append or unique in a bottom up manner, one can convert the expression into an equivalent translatable one.

The equivalences in Thm. 1 can be used to prove that Trans is semantics preserving. For example, for $\text{Trans}(\sigma(s, f_{\sigma_2}))$:

$$
\begin{aligned}
\sigma(s, f_{\sigma_2}) &= \sigma(\pi(\text{sort}(\sigma(b, f_\sigma), l_{sort}), f_\pi), f_{\sigma_2}) \\
&= \pi(\sigma(\text{sort}(\sigma(b, f_\sigma), l_{sort}), f_{\sigma_2}), f_\pi) \\
&= \pi(\text{sort}(\sigma(b, f_\sigma \wedge f_{\sigma_2}), l_{sort}), f_\pi) \\
&= \text{Trans}(\sigma(s, f_{\sigma_2}))
\end{aligned}
$$

$\square$

Note that in the definition of Trans, a sort with an empty list of fields is simply the identity function and has no effect on the order of the list. This is relevant because in the cases where the argument to Trans is already a baseExp, Trans simply wraps this argument with dummy selects, sorts and projections to get a translatable expression.

Let $s_i = \pi(\text{sort}(\sigma(b, f_{\sigma_i}), l_{sort_i}), f_{\pi_i})$. Trans is defined on expressions whose subexpressions (if any) are in translatable form, so we have to consider cases where the sub-expressions are either $s$ or $\text{top}(s)$. Each case is defined below.

$\boxed{\text{Query}(...)}$

$\text{Trans}(\text{Query}(...))) = \pi(\text{sort}(\sigma(\text{Query}(...), \text{True}), [\,]), f)$

where $f$ projects all the fields from the input relation.

$\boxed{\pi(t, f_{\pi_2})}$

$\text{Trans}(\pi(s, f_{\pi_2})) = \pi(\text{sort}(\sigma(b, f_\sigma), l_{sort}), f')$
$\text{Trans}(\pi(\text{top}(s, e), f_{\pi_2})) = \text{top}(\pi(\text{sort}(\sigma(b, f_\sigma), l_{sort}), f'), e)$

where $f'$ is the composition of $f_\pi$ and $f_{\pi_2}$.

$\boxed{\sigma(t, f_{\sigma_2})}$

$\text{Trans}(\sigma(s, f_{\sigma_2})) = \pi(\text{sort}(\sigma(b, f_\sigma \wedge f_{\sigma_2}), l_{sort}), f_\pi)$
$\text{Trans}(\sigma(\text{top}(s, e), f_\sigma)) = \text{top}(\pi(\text{sort}(\sigma(b, f_\sigma \wedge f_{\sigma_2}), l_{sort}), f_\pi), e)$

$\boxed{\bowtie (t_1, t_2, f_\bowtie)}$

$\text{Trans}(\bowtie (s_1, s_2, f_\bowtie))$
$= \pi(\text{sort}(\sigma(\bowtie (b_1, b_2, \text{True}), f'_\sigma), l'_{sort}), f'_\pi)$

where $f'_\sigma = f_{\sigma_1} \wedge f_{\sigma_2} \wedge f_\bowtie$, $l'_{sort} = l_{sort_1} : l_{sort_2}$, and $f'_\pi$ is the concatenation of the fields from $f_{\pi_1}$ and $f_{\pi_2}$.

$\text{Trans}(\bowtie (\text{top}(s_1, e), \text{top}(s_2, e)))$
$= \pi(\text{sort}(\sigma(\bowtie (\text{top}(s_1, e), \text{top}(s_2, e)), \text{True}), [\,]), \text{id})$

where id is the identity mapping from input to output record.

$\boxed{\text{top}(t, e)}$

$\text{Trans}(\text{top}(s, e)) = \text{top}(s, e)$
$\text{Trans}(\text{top}(\text{top}(s, e_1), e_2)) = \text{top}(s, e')$

where $e'$ is the maximum value of $e_1$ and $e_2$.

$\boxed{\text{agg}(t)}$

$\text{Trans}(\text{agg}(s)) = \pi(\text{sort}(\sigma(\text{agg}(s), \text{True}), [\,]), \text{id})$
$\text{Trans}(\text{agg}(\text{top}(s, e))) = \pi(\text{sort}(\sigma(\text{agg}(s), \text{True}), [\,]), \text{id})$

where id is the identity mapping from input to output record.

$\boxed{\text{sort}(t)}$

$\text{Trans}(\text{sort}(s, l_{sort})) = \pi(\text{sort}(\sigma(b, f_\sigma), l'_{sort}), f_\pi)$
$\text{Trans}(\text{sort}(\text{top}(s, e))) = \text{top}(\pi(\text{sort}(\sigma(b, f), l'_{sort}), f_\pi), e)$

where $l'_{sort} = l_s : l_{sort}$.

**Figure 5.** Definition of Trans

**Translatable expressions to SQL.** Following the syntax-directed rules in Fig. 6, it is possible to convert any translatable expression into an equivalent SQL expression. Most of the rules in Fig. 6 are direct translations from the operators in the theory into their SQL equivalents.

One important aspect of the translation is the way that ordering of records is preserved. Ordering is problematic because although the operators in the theory define the order of the output in terms of the order of their inputs, SQL queries are not guaranteed to preserve the order of records from nested sub-queries; e.g., the ordering imposed by an ORDER BY clause in a nested query is not guaranteed to be respected by an outer query that does not impose any ordering on the records.

To solve this problem, the translation rules introduce a function getOrder, which scans a translatable expression $t$ and returns a list

$$
\begin{aligned}
[\![\mathrm{Query}(string)]\!] &= (\ string\ ) \\
[\![\mathrm{top}(s, e)]\!] &= \mathrm{SELECT}\ *\ \mathrm{FROM}\ [\![s]\!]\ \mathrm{LIMIT}\ [\![e]\!] \\
[\![\bowtie (t_1, t_2, \mathrm{True})]\!] &= \mathrm{SELECT}\ *\ \mathrm{FROM}\ [\![t_1]\!],\ [\![t_2]\!] \\
[\![\mathrm{agg}(t)]\!] &= \mathrm{SELECT}\ \mathrm{agg}(field)\ \mathrm{FROM}\ [\![t]\!] \\
[\![\pi(\mathrm{sort}(\sigma(t, f_\sigma), e), f_\pi)]\!] &= \mathrm{SELECT}\ [\![f_\pi]\!]\ \mathrm{FROM}\ [\![t]\!]\ \mathrm{WHERE}\ [\![f_\sigma]\!] \\
&\quad \mathrm{ORDER\ BY}\ [\![e]\!], \mathrm{getOrder}(t) \\
[\![\mathrm{unique}(t)]\!] &= \mathrm{SELECT\ DISTINCT}\ *\ \mathrm{FROM}\ [\![t]\!] \\
&\quad \mathrm{ORDER\ BY\ getOrder}(t) \\[6pt]
[\![f_\sigma(e)]\!] &= [\![e]\!].f_1\ op\ [\![e]\!]\ \mathrm{AND}\ \ldots\ [\![e]\!].f_N\ op\ [\![e]\!] \\
[\![f_\sigma(e, t)]\!] &= [\![e]\!]\ \mathrm{IN}\ [\![t]\!] \\
[\![f_\pi(e)]\!] &= [\![e]\!].f_1, \ldots, [\![e]\!].f_N
\end{aligned}
$$

**Figure 6.** Syntax-driven rules for SQL translation from translatable expressions

$$
\begin{aligned}
\mathrm{getOrder}(\mathrm{Query}(\ldots)) &= [\,] \\
\mathrm{getOrder}(\mathrm{top}(e, i)) &= \mathrm{getOrder}(e) \\
\mathrm{getOrder}(\pi(e, f)) &= \mathrm{getOrder}(e) \\
\mathrm{getOrder}(\sigma(e, f)) &= \mathrm{getOrder}(e) \\
\mathrm{getOrder}(\bowtie (e_1, e_2, f)) &= \mathrm{getOrder}(e_1) : \mathrm{getOrder}(e_2) \\
\mathrm{getOrder}(\mathrm{sort}(e, [e.f_i])) &= [e.f_i] : \mathrm{getOrder}(e) \\
\mathrm{getOrder}(\mathrm{unique}(e)) &= \mathrm{getOrder}(e) \\
\mathrm{getOrder}(\mathrm{agg}(e)) &= [\,]
\end{aligned}
$$

**Figure 7.** Definition of getOrder

of fields that are used to order the subexpressions in $t$. The list is then used to impose an ordering on the outer SQL query with an ORDER BY clause. The function is defined for each of the operators in the theory in Fig. 7. One detail of the algorithm not shown in the figure is that some projections in the inner queries need to be modified so they do not eliminate fields that will be needed by the outer ORDER BY clause.

**Append and Unique.** The append operation is not included in translatable expressions because there is no simple means to combine two relations in SQL that preserves the ordering of records in the resulting relation. We can still translate unique, however, using the SELECT DISTINCT construct at the outermost level, as Fig. 6 shows. Using unique in nested expressions, however, can change the semantics of the results in ways that are difficult to reason about (e.g., unique(top($r, e$)) is not equivalent to top(unique($r$), $e$)). Thus, the only expressions with unique that we translate to SQL are those that use it at the outermost level. In our experiments, we found that omitting those two operators did not significantly limit the expressiveness of the theory.

## 4. Language for Persistent Data Manipulation

Having presented the theory of ordered relations, we now describe how we use it to automatically convert imperative code to SQL expressions. We begin by describing the modeling of the input program using a kernel language. The kernel language allows us to define simple rules to compute verification conditions as will be shown in subsequent sections.

### 4.1 A Kernel Language

We use a simple imperative language to model Java programs that retrieve values from a database and compute on them. The language operates on three types of values: scalars, immutable records,

$$
\begin{aligned}
c \in \mathrm{constant} &::= \mathrm{True} \mid \mathrm{False} \mid \text{number literal} \mid \text{string literal} \\
e \in \mathrm{expression} &::= c \mid var \mid e.f \mid \{f_i = e_i\} \mid e_1\ op\ e_2 \mid \neg\ e \\
&\quad \mid [\,] \mid \mathrm{Query}(\ldots) \mid \mathrm{size}(e) \mid \mathrm{get}(e_r, e_s) \\
&\quad \mid \mathrm{append}(e_1, e_2) \mid \mathrm{unique}(e) \\
c \in \mathrm{command} &::= \mathrm{skip} \mid var := e \mid \mathrm{if}(e)\ \mathrm{then}\ c_1\ \mathrm{else}\ c_2 \\
&\quad \mid \mathrm{while}(e)\ \mathrm{do}\ c \mid c_1\ ;\ c_2 \mid \mathrm{assert}\ e \\
op \in \mathrm{binary\ op} &::= \wedge \mid \vee \mid > \mid < \mid \geq \mid \leq \mid =
\end{aligned}
$$

**Figure 8.** Abstract syntax of the kernel language

and immutable lists. Lists are used to represent the collection of records and are modeled after the Iterable interface in Java. They are also used to model the results that are returned from database retrieval operations. Lists store either scalar values or records constructed with scalars, and nested lists are assumed to be appropriately flattened. The language currently does not model the three-valued logic of null values in SQL, and does not model updates to the database. We assume that there is a standard type system in place to ensure that the input program is well-typed. Figure 8 shows the syntax of the language.

The semantics of the constructs in the kernel language are mostly standard, with a few new ones introduced for record retrievals. [ ] constructs creates an empty list. Query(...) retrieves records from the database and the results are returned as a list. The records of a list can be randomly accessed using get, and records can be appended to a list using append. Finally, unique takes in a list and creates a new list with all duplicate records removed.

### 4.2 Conversion from Java Programs

Unfortunately, real-world programs are not written in the kernel language. Most of the primitive Java operations can be converted easily after standard transformations such as expression flattening and function inlining. We currently do not handle polymorphic types and exceptions. In the following, we describe two types of transformations from Java programs into the kernel language: the first group of transformations expand the expressiveness of the language, and another group for preserving Java semantics.

**Java list operations.** The Java collections classes contain many methods on lists that are not modeled by the kernel language. To handle that issue, we implement a number of such operations using the kernel language (e.g., contains, clear, iterators, etc). Doing so allows us to process more real-world programs as our experiments show.

**Java data structures.** The kernel language assumes that the user program uses lists as the data structure to manipulate persistent data, but Java includes other data structures as well. While the language does not directly support other data structures, we can, however, implement them using lists. For instance, we model hashtable as a list of (key, value) pairs, with value being a flattened list of records that are mapped to the same key. Inserting into the hashtable is simply appending the appropriate list of values, and retrieval is implemented by iterating through the list until the given key is found. Modeling hashtables allow our prototype to recognize a broader set of relational operations in imperative code (such as hash join), and other data structures can be similarly modeled as well.

**Boolean variables.** The kernel language includes boolean values but not boolean variables. As such, boolean variables in the input source are converted into integers, and boolean operations are converted into integer operations. This allows us to process more classes of code fragments, as our experiments show.

$$\begin{aligned}
\text{VC}(\text{skip}, P) &= P \\
\text{VC}(lv := e, P) &= P[e/lv] \\
\text{VC}(\text{if}(e) \text{ then } c_1 \text{ else } c_2, P) &= (e \wedge \text{VC}(c_1, P)) \vee (\neg e \wedge \text{VC}(c_2, P)) \\
\text{VC}(\text{while}(e) \text{ do } c, P) &= I \wedge \forall x_1, ..., x_N I \to (e \to \text{VC}(c, I) \\
&\quad \wedge \neg e \to P \\
\text{VC}(c_1 \, ; \, c_2, P) &= \text{VC}(c_1, \text{VC}(c_2, P)) \\
\text{VC}(\text{assert } e, P) &= P \wedge e
\end{aligned}$$

**Figure 9.** Rules for computing verification conditions for the kernel language

**Statements with side effects.** The input source code might contain Java statements with side effects while iterating through records retrieved from the database. Our kernel language does not model side effects. In order to process such loops, we remove statements with side effects from the loop and attempt to convert the remainder of the loop into SQL. If our algorithm is able to convert the loop with a SQL query, we create a new loop with the statements with side effects as the body, and have the loop iterate over the results from the converted SQL query instead.

**Preserving Java semantics on lists.** The kernel language assumes that the list operations do not raise errors or exceptions, which is not the case for Java programs. For instance, Java get throws an exception when the index is out of bounds. To model such behavior, we prepend all $\text{get}(r, e)$ with $\text{assert } e < \text{size}(r)$. The added assertion is incorporated into the verification condition. As a result, if the validity of the prepended assertion cannot be established then QBS will not be able to convert the input program into SQL.

## 5. Generating Verification Conditions

Given an input program in the kernel language, the next step in the our algorithm is to come up with an expression for the postcondition variable of the form $pcVar = e$, where $e$ is a translatable expression as defined in Sec. 3.4. In order to infer the postcondition, we first compute *verification conditions* over each line of code in the input program. These conditions amount to assertions that must hold true over the input program, and are expressed using the predicate language as presented in Fig. 3.

To compute verification conditions, we employ standard results from axiomatic semantics [19] in systematic verification of program behavior. The rules that are used to compute verification conditions for the commands in the kernel language are shown in Fig. 9.

The statement $\text{VC}(c, P) = P'$ means that given postcondition $P$, the verification condition for command $c$ is $P'$. Computation of verification condition is relatively straight-forward. For instance, to compute the verification condition for the assignment command $lv := e$ given postcondition $P$, we simply replace occurrences of $lv$ in $P$ with $e$. As in traditional Hoare logic style verification, computing the verification condition of the while statements involves a loop invariant $I$ and states that: 1. the invariant has to be true prior to entry into the loop; 2. if the loop condition $e$ is true, and that invariant is true for all the program variables $x_1, ..., x_N$ that are modified in the loop body $c$, then the verification condition of the loop body given the invariant as the postcondition is true (i.e., the loop is preserved) and; 3. when the loop terminates, the postcondition that we would like to establish, $P$, is true.

Unlike traditional computation of verification conditions, both the postcondition and the loop invariants are unknown when the conditions are generated. This does not pose problems for our algorithm as we simply treat invariants (and the postcondition) as functions of the program variables that are currently in scope when the loop is entered, and computation of verification conditions proceeds as described.

| Verification conditions for the outer loop | |
|---|---|
| initialization | outerLoopInvariant(0, *users*, *roles*, [ ]) |
| termination | $i \geq \text{size}(users) \wedge \text{outerLoopInvariant}(i, users, roles, listUsers) \to$ postcondition(*listUsers*, *users*, *roles*) |
| perservation | (same as inner loop initialization) |
| **Verification conditions for the inner loop** | |
| initialization | $i < \text{size}(users) \wedge \text{outerLoopInvariant}(i, users, roles, listUsers) \to$ innerLoopInvariant(*i*, 0, *users*, *roles*, *listUsers*) |
| termination | $j \geq \text{size}(roles) \wedge \text{innerLoopInvariant}(i, j, users, roles, listUsers) \to$ outerLoopInvariant(*i* + 1, *users*, *roles*, *listUsers*) |
| preservation | $j < \text{size}(roles) \wedge \text{innerLoopInvariant}(i, j, users, roles, listUsers) \to$ $(\text{get}(users, i).id = \text{get}(roles, j).id \wedge$ innerLoopInvariant(*i*, *j* + 1, *users*, *roles*, append(*listUsers*, get(*users*, *i*)))) $\vee$ $(\text{get}(users, i).id \neq \text{get}(roles, j).id \wedge$ innerLoopInvariant(*i*, *j* + 1, *users*, *roles*, *listUsers*)) |

**Figure 10.** Verification conditions for the running example

As an example, Fig. 10 shows the verification conditions that are generated for the running example. In this case, the verification conditions are split into two parts, with invariants defined for both loops. The first two assertions describe the behavior of the outer loop on line 5, with the first one asserting that the outer loop invariant must be true on entry of the loop (after applying the rule for the assignments prior to loop entry), and the second one asserting that the postcondition for the loop is true when the loop terminates. The third assertion asserts that the inner loop invariant is true when it is first entered, given that the outer loop condition and loop invariant are true. The preservation assertion is the inductive argument that the inner loop invariant is preserved after executing one iteration of the loop body. The list *listUsers* is either appended with a record from get(*users*, *i*), or is remain unchanged, depending on whether the condition for the if statement, get(*users*, *i*).*id* = get(*roles*, *j*).*id*, is true or not. Finally, the termination assertion states that the outer loop invariant is valid when the inner loop terminates.

## 6. Synthesis of Invariants and Postconditions

The verification conditions computed from the previous section lack definitions for loop invariants and postconditions. We synthesize these predicates using the SKETCH constraint-based synthesis system [28]. SKETCH uses a counterexample guided synthesis algorithm (CEGIS) to efficiently search very large spaces of candidate expressions for one that is correct according to a bounded model checking procedure. In order to make the search tractable, SKETCH takes as input a definition of the space of expressions to search. In this section, we show how the QBS algorithm analyzes the input program to infer a space of candidate expressions for the synthesizer to search. QBS actually generates these spaces incrementally, so if the synthesizer fails to find a solution in a smaller space, a bigger space is attempted.

$$i\ op\ \left\{\begin{array}{c} i \mid \mathsf{size}(\mathit{users}) \mid \mathsf{size}(\mathit{roles}) \mid \mathsf{size}(\mathit{listUsers}) \mid \\ \mathsf{sum}(\pi(\mathit{users}, f)) \mid \mathsf{sum}(\pi(\mathit{roles}, f)) \mid \mathsf{max}(\pi(\mathit{users}, f)) \mid \\ \ldots \end{array}\right\} \wedge \mathit{listUsers}\ op\ \left\{\begin{array}{c} \mathit{listUsers} \mid \sigma(\mathit{users}, f) \mid \\ \bowtie (\mathsf{top}(\mathit{users}, e_1), \mathsf{top}(\mathit{listUsers}, e_2), f) \mid \\ \bowtie (\sigma(\mathsf{top}(\mathit{users}, e_1), f_1), \sigma(\mathsf{top}(\mathit{listUsers}, e_2), f_2), f_3) \mid \\ \ldots \end{array}\right\}$$

**Figure 11.** Space of possible invariants for the outer loop of the running example.

### 6.1 Inferring the Space of Possible Invariants

Recall that each invariant is parameterized by the current program variables that are in scope. Our algorithm assumes that each loop invariant is a conjunction of predicates, with each predicate having the form $lv_i = e$, where $lv$ is a program variable that is modified within the loop, and $e$ is an expression from the predicate grammar in Fig. 3.

The space of expressions $e$ is restricted to expressions of the same static type as $lv$ involving the variables that are in scope. The system limits the size of expressions that the synthesizer can consider, and incrementally increases this limit if the synthesizer fails to find any candidate solutions (to be explained in Sec. 6.3).

Fig. 11 shows the set of candidate predicates for the outer loop in the running example. The figure shows the potential expressions for the program variable $i$ and $\mathit{listUsers}$. The definitions for the various functions used in the relational operators are generated in a similar way.

One advantage of using the theory of ordered relations is that invariants can be relatively concise. This has a big impact for synthesis, because the space of expressions grows exponentially every time the size of expressions to search over increases.

### 6.2 Creating Templates for Postconditions

The process of creating templates for postconditions is similar to that of creating templates for invariants. The difference is that we only need to generate a template for the postcondition variable, since we are not seeking to generate the strongest postconditions. The mechanism used to generate possible expressions for the postcondition variable is similar to that for invariants, except that:

• the postcondition must be in the form $pcVar = e$ and not any other binary operators between $pcVar$ and $e$, as those do not lead to meaningful SQL translations.

• similarly, we do not generate the identity form, as that is trivially true and is not a meaningful expression.

With that in mind, the algorithm would generate the following template for the postcondition of the running example:

$$\mathit{listUsers} = \left\{\begin{array}{c} \mathit{users} \mid \sigma(\mathit{users}, f) \mid \mathsf{top}(\mathit{users}, e) \mid \\ \bowtie (\mathsf{top}(\mathit{users}, e_1), \mathsf{top}(\mathit{roles}, e_2), f) \mid \\ \bowtie (\sigma(\mathsf{top}(\mathit{users}, e_1), f_1), \sigma(\mathsf{top}(\mathit{roles}, e_2), f_2), f) \mid \\ \ldots \end{array}\right\}$$

Again, the functions and expressions used in the relational operations generated as above.

### 6.3 Optimizations

The basic algorithm presented above for generating invariant and postcondition templates is sufficient but not efficient for synthesis. In this section we describe two optimizations that improve the synthesis efficiency.

**Incremental solving.** As an optimization, the generation of templates for invariants and postconditions is done in an iterative manner: our algorithm initially creates simple templates using the production rules from the predicate grammar, such as considering expressions with only one relational operator, and functions that contains only one boolean clause. If the synthesizer is able to generate a candidate that is sufficiently strong to confirm the validity of the verification conditions, then our job is done. Otherwise, the algorithm repeats the template generation process, but increases the complexity of the template that is generated by considering expressions consisting of more relational operators, and more complicated boolean functions. Our evaluation using real-world examples shows that most code examples require only a few ($< 5$) iterations before finding a candidate solution. Additionally, the incremental solving process can be done in a parallel fashion (although we did not implement this), further decreasing runtime if needed.

**Breaking symmetries.** Symmetries have been shown to be one of sources of inefficiency in constraint solvers [12, 31]. Unfortunately, the template generation algorithm presented above can generate highly symmetrical expressions. For instance, it can generate the following potential candidates for the postcondition:

$$\sigma(\sigma(\mathit{users}, f_1), f_2)$$
$$\sigma(\sigma(\mathit{users}, f_2), f_1)$$

Notice that the two expressions are semantically equivalent to the expression $\sigma(\mathit{users}, f_1 \wedge f_2)$. However, the synthesizer does not have that knowledge, and would consider the two expressions individually, leading to inefficiency. Fortunately, the equivalence theorem developed in Sec. 3.3 allows us to prune away a large subset of symmetrical expressions. The template generation algorithm uses the theorem as much as possible during expression generation to reduce the number of expressions that are actually created for both invariants and postconditions.

One further optimization is applied when generating templates for postconditions. Since our goal is to translate the postcondition into SQL, we only need to generate translatable expressions as defined in Sec. 3.4 as potential candidates. Our experiments have shown that applying these symmetric breaking optimizations can reduce the amount of solving time by half.

The current template generation algorithm is simple in that for a given variable $lv$, it only uses scope and static types as means to restrict the number of other variables that can be potentially related to $lv$ for invariant and postcondition purposes. More static analysis can be performed on the source code to further restrict the number of potential variables, for instance using dataflow analysis to detect variable relationships. However, given that our current prototype can already infer a variety of imperative constructs in SQL queries within a reasonable amount of time, we decided to leave such extensions for future work.

## 7. Formal Validation and Source Transformation

After the synthesizer comes up with candidate invariants and postconditions, they need to be validated using a theorem prover, since the synthesizer used in our prototype is only able to perform bounded reasoning as discussed earlier. We have implemented the theory of ordered relations in the Z3 [5] prover for this purpose. Since the theory of lists is not decidable as it uses universal quantifiers, the theory of ordered relations is not decidable as well. However, for practical purposes we have not found that to be limiting in our experiments. In fact, given the appropriate invariants and postconditions, the prover is able to validate them within seconds by making use of the axioms that are provided.

If the prover can establish the validity of the invariants and postcondition candidates, the postcondition is then converted into SQL according to the rules discussed in Sec. 3.4. For instance, for the

8     *2012/8/10*

| Type | Expression inferred |
|---|---|
| outer loop invariant | $i \leq \text{size}(users) \land$ $listUsers = \pi(\text{sort}(\sigma(\bowtie (\text{top}(users, i), roles, \text{True}), f_\sigma), l_{sort}, f_\pi)$ |
| inner loop invariant | $i < \text{size}(users) \land j \leq \text{size}(roles) \land$ $listUsers = \text{append}($ $\pi(\text{sort}(\sigma(\bowtie (\text{top}(users, i), roles, \text{True}), f_\sigma), l_{sort}), f_\pi),$ $\pi(\text{sort}(\sigma(\bowtie (\text{get}(users, i), \text{top}(roles, j), \text{True}), f_\sigma), l_{sort}), f_\pi))$ |
| postcondition | $listUsers = \pi(\text{sort}(\sigma(\bowtie (users, roles, \text{True}), f_\sigma), l_{sort}, f_\pi)$ |

where
$f_\sigma := \text{get}(users, i).roleId = \text{get}(roles, j).roleId$
$f_\pi$ projects all the fields from the *User* class
$l_{sort} = [users, roles]$

**Figure 12.** Inferred expressions for the running example

```
List objs1 = fetchRecordsFromDB();
List results1 = new ArrayList();
for (Object o : objs1) {
  if (f(o))
    results1.add(o);
}
List results2 = new ArrayList();
for (Object o : objs1) {
  if (g(o))
    results2.add(o);
}
```

**Figure 13.** Code fragment with alias in results

running example our algorithm found the invariants and postcondition as shown in Fig. 12, and the input code is transformed into the results in Fig. 2.

If the prover is unable to establish validity of the candidates (detected via a timeout), we ask the synthesizer to generate other candidate invariants and postconditions after increasing the space of possible solutions as described in Sec. 6.3. One reason that the prover may not be able to establish validity is because the maximum size of the relations set for the synthesizer was not large enough. For instance, if the code returns the first 100 elements from the relation but the synthesizer only considers relations up to size 10, then it will incorrectly generate candidates that claim that the code was performing a full selection of the entire relation. In such cases our algorithm will repeat the synthesis process after increasing the maximum relation size.

### 7.1 Object Aliases

Implementations of ORM libraries typically create new objects from the records that are fetched, and our current implementation will only transform the input source into SQL if all the objects involved in the code fragment are freshly fetched from the database, as in the running example. In some cases this may not be true, as in the code fragment in Fig. 13.

Here, the final contents of results1 and results2 can be aliases to those in objs1. In that case, rewriting results1 and results2 into two SQL queries with freshly created objects will not preserve the alias relationships in the original code. Our current implementation will not transform the code fragment in that case, and we leave sharing record results among multiple queries as future work.

## 8. Experiments

In this section we report our experiment results. The goal of the experiments is twofold: first, to quantify the ability of our algorithm to convert Java code into real-world applications, and second to explore the limitations of the current implementation.

We implemented a prototype of the QBS algorithm. The source code analysis and computation of verification conditions are implemented using the Polyglot compiler framework [25]. We use Sketch [28] as the synthesizer for invariants and postconditions, and Z3 [5] for validating the invariants and postconditions.

### 8.1 Real-World Benchmarks

In the first set of experiments, we evaluated our implementation of the QBS algorithm using real-world examples from two large-scale open-source applications written in Java. Both applications use the Hibernate ORM library for data persistence operations. Wilos [4] is a project management application with 62k LOC, and itracker [1] is a software issue management system with 61k LOC. We randomly selected a number of classes that are stored persistently and manually identified code fragments that fetch multiple instances of such classes from the database and manipulate them in loop constructs (such as the running example in Fig. 1), or pass them to methods from Java's collection classes (such as Collections.sort). We removed fragments that we believed would not benefit from query inference (such as those that iterate through all fetched objects from the base query and mutate each one). This process resulted in a total of 49 code fragments. We pass each fragment to our QBS prototype. Figure 14 reports the number of fragments that QBS was able to transform into SQL equivalent expressions. For comparison purposes, we also report the number of benchmarks that we believe can be transformed by prior work [32, 33], based on the published description of these algorithms, where relational projections and selections are inferred by computing access paths in loop constructs. The numbers were inferred by manual inspecting each benchmark and checking if it performs a selection or projection in imperative code. The details of the benchmarks are given in the appendix.

| Application | # benchmarks | # trans. by prior work | # trans. by QBS |
|---|---|---|---|
| Wilos | 33 | 9 | 28 |
| itracker | 16 | 5 | 12 |
| **Total** | **49** | **14** | **40** |

**Figure 14.** Experiment on real-world benchmarks

### 8.2 Discussion

First of all, the experiment shows that the QBS algorithm is able to infer relational specifications from a large number of real-world benchmarks and convert them into SQL equivalents. For the benchmarks that are reported as translatable by QBS, our prototype was able to synthesize postconditions and invariants, and also validate them using the prover. Furthermore, our prototype was able to process most of benchmarks under 5 minutes, with the ones that involve join operations consuming the most amount of time (up to 40 mins maximum). In the following, we broadly describe the common types of relational operations that our QBS prototype inferred from the benchmarks, along with some limitations of the current implementation.

**Projections and Selections.** A number of benchmarks perform relational projections and selections in imperative code. Typical projections include selecting specific fields from the list of records that are fetched from the database, and selections include filtering a subset of objects using field values from each object (e.g., user ID equals to some numerical constant), and a few use criteria that involve program variables that are passed into the method.

One special case is worth mentioning. In some cases only a single field is projected out and loaded into a set data structure,

9                                                                                                                              2012/8/10

such as a set of integer values. One way to translate such cases (for instance, using the scheme from prior work) is to generate SQL that fetches the field from all the records (including duplicates) into a list, and subsequently eliminate the duplicates and return the set to the user code. Our prototype, however, improves upon that scheme by detecting the type of the postcondition variable and inferring a postcondition involving the unique operator, which is then translated to a SELECT DISTINCT query that avoids fetching duplicate records from the database.

**Joins.** Another set of benchmarks involve join operations. We summarize the join operations in the application code into two categories. The first involves obtaining two lists of objects from two base queries and looping through each pair of objects in a nested for or while loop. The pairs are filtered and (typically) one of the objects from each pair is retained. The running example in Fig. 1 represents such a case. For these cases, our prototype translates the program fragment into a relational join of the two base queries with the appropriate join predicate, projection list, and sort operations that preserve the ordering of records in the results.

Another type of join also involves obtaining two lists of objects from two base queries. Instead of a nested loop join, however, the code iterates through each object e from the first list, and searches if e (or one of e's fields) is contained in the second list. If the result is true, then e (or some of its fields) is appended to the resulting list and is returned at the end. For these cases our prototype converts the search operation into a contains expression in the predicate language, after which the expression is translated into a correlated subquery in the form of SELECT * FROM r1, r2 WHERE r1 IN r2, with r1 and r2 being the base queries.

As mentioned, our prototype is able to handle both join idioms mentioned above. However, the loop invariants and postconditions involved in such cases tend to be more complex as compared to selections and projections, as illustrated by the running example in Fig. 12. As a result, they require more iterations of the synthesis and formal validation before a valid solution can be found, with up to 40 mins in the longest case and the majority of the time spent in synthesis and bounded verification. Unfortunately, we are not aware of any prior techniques that can be used to infer join queries from imperative code, and we believe that more optimizations can be devised to speed up the synthesis process for such cases.

**Aggregations.** Aggregations are used in benchmarks in a different number of ways. The most straightforward ones are those that simply return the length of the list that is returned from an ORM query, which are translated into COUNT queries. More sophisticated uses of aggregates include iterating through all records in a list to find the maximum or minimum values, or searching if a record exists in a list. Aggregates such as maximum and minimum are interesting as they introduce loop-carried dependencies [6], where the running value of the aggregate is updated conditionally based on the value of the current record as compared to previous ones. By using the top operator from the theory of ordered relations, our prototype is able to generate a loop invariant of the form v = agg(top(r, i)) and subsequently translate the postcondition into the appropriate SQL query.

As a special case, a number of benchmarks check for the existence of a particular record in a relation by iterating over all records and setting a result boolean variable to be true if it exists. As mentioned in Sec. 4.2, boolean variables and operations on booleans are converted into integer equivalents. Generating invariants that are similar as in the case of the other aggregates, our prototype translates such code fragments into SELECT COUNT(*) > 0 FROM ... WHERE e, where e is the expression to check for existence in the relation. We rely on the database query optimizer to further rewrite this query into the more efficient form using EXISTS.

**Limitations.** Finally, there are a few examples from the two applications where our prototype fails to translate the code fragment into SQL, even though we believe that there is an equivalent SQL query. For instance, some benchmarks include code fragment where the input list is sorted by calling Collections.sort, followed by retrieving the last record from the sorted list, which is equivalent to max or min depending on the sort order. Including extra axioms in the theory would allow us to reason about such cases. Another set of benchmarks include advanced use of types, such as storing polymorphic records in the database, and performing different operations based on the type of records retrieved. Incorporating type information in the theory of ordered relations is an interesting area for future work.

### 8.3 Advanced Idioms

In the second part of experiments, we used synthetic code fragments to demonstrate the ability of our prototype to translate more complex expressions into SQL. Although we did not find such examples in either of our two real-world applications, we believe that these can occur in real applications.

**Hash Joins.** Beyond the join operations that we encountered in the application benchmarks, we wrote two synthetic test cases for joins that join relations r and s using the predicate r.a = s.b, where a and b are integer fields. In the first case, the join is done via hashing, where we first iterate through records in r and build a hashtable, whose keys are the values of the a field, and where each key maps to a list of records from r that has that corresponding value of a. We then loop through each record in s to find the relevant records from r to join with, using the b field as the look up key. As mentioned in Sec. 4.2, the QBS algorithm models hashtables using lists, and with that our prototype is able recognize this process as a join operation and convert the fragment accordingly, similar to the join benchmarks mentioned above.

**Sort-Merge Joins.** Our second synthetic test case joins two lists by first sorting r and s on fields a and b respectively, and then iterating through both lists simultaneously. We advance the scan of r as long as the current record from r is less than (in terms of fields a and b) the current record from s, and similarly advance the scan of s as long as the current s record is less than the current r record. Records that represent the join results are created when the current record from r equals to that from s on the respective fields. Unfortunately, our current prototype fails to translate the code fragment into SQL, as the invariants for the loop cannot be expressed using the current the predicate language, since that involves expressing the relationship between the current record from r and s with all the records that have been previously processed.

**Iterating over Sorted Relations.** We next tested our prototype with two usages of sorted lists. We created a relation with one unsigned integer field id as primary key, and sorted the list using the sort method from Java. We subsequently scanned through the sorted list as follows:

```
List records = Query("SELECT id FROM t");
List results = new ArrayList();
Collections.sort(records); // sort by id
for (int i = 0; i < 10; ++i)
  results.add(records.get(i));
```

Our prototype correctly processes this code fragment by translating it into SELECT id FROM t ORDER BY id LIMIT 10. However, if the loop is instead written as follows:

```
List records = Query("SELECT id FROM t");
List results = new ArrayList();
Collections.sort(records); // sort by id
int i = 0;
while (records.get(i).id < 10) {
  results.add(records.get(i));
  ++i;
}
```

The two loops are equivalent since the id field is a primary key of the relation, and thus there can at most be 10 records retrieved. However, our prototype is not able to reason about the second code fragment, as that requires an understanding of the schema of the relation, and that iterating over id in this case is equivalent to iterating over i in the first code fragment. Both of which require additional axioms to be added to the theory before such cases can be converted.

## 9. Related Work

The idea of inferring relational specifications from imperative code constructs was first studied in [32, 33]. The work uses abstract interpretation and attribute grammar to extract SQL queries from Java programs. The idea is to compute the set of data access paths that a piece of imperative code traverses, and replace the imperative code that performs the explicit path traversal with SQL queries. The analysis can be applied to recursive function calls, but does not handle code fragments with loop-carried dependencies, or fragments that implement join or aggregation. It is unclear how this method (modeling relational operations as access paths) can be extended to handle such cases.

In contrast, our implementation is able to infer both relational join and aggregation from imperative code. We currently do not handle recursive function calls, although we have not encountered the use of recursive functions in the two real-world applications used in our experiments.

**Modeling relational operations.** Our ability to infer relational specifications from imperative code relies on using the theory of ordered relations to connect the imperative and relational worlds. Previous work modeled such operations differently. In addition to modeling relational operations using data access paths, other work has modeled relational databases using bags [10], sets [22], and nested relational calculus [34]. In that respect, one key insight of our work is that ordered relations are the right abstraction for our purpose, as they are similar to the interfaces provided by the ORM libraries in the imperative code, and allow us to design a sound and precise transformation into SQL. Other model choices, such as the theory of sets, would be closer to the relational model, but would make reasoning about operations such as top difficult, as elements in a set are not ordered.

To our knowledge, our work is the first one to address the issue of precision, i.e., the ordering of records, in relational transformations. Precision would not be an issue if the source program only operated on orderless data structures such as sets or did not perform any relational operations that create new records, such as joins. Unfortunately, most imperative languages (such as Java) provide interfaces based on ordered data structures, making this problematic. The benchmarks in our experiments also show that implementing joins in imperative code is a relatively common operation.

**Automatic mining of program invariants.** Our verification-based approach to finding a translatable postcondition is similar to that used in earlier work [20, 21], although previous work acknowledges that finding invariants is difficult and instead uses a symbolic execution based approach in reasoning about program behavior in loops. The idea of scanning the source program to generate the synthesis template is inspired by the PINS algorithm [30], although in our work we do not require user intervention. Meanwhile, there has been substantial earlier work on automatically detecting loop invariants, such as using predicate refinement [15] or dynamic approaches like the Daikon system [14] and invGen [18]. There has also been work on using synthesis to discover invariants [29]. Our work differs from previous approaches in that we only need to discover invariants and postconditions that are *strong enough* to validate the transformation, and our predicate grammar and translatable expressions greatly prune the space of invariants to those needed for common relational operations, rather than those needed for general-purpose programs.

**Integrated query languages.** Integrating programming languages and database query languages into a single language has been an active research area, with projects such as LINQ [24], Kleisli [34], Links [11], JReq [20], the functional language proposed in [10], Ferry [17], and DBPL [27]. These solutions provide a means to embed database queries in imperative programs without using ORM-like libraries, or requiring users to embed SQL in their code. Unfortunately, many of them do not support all relational operations, and the syntax of many of such languages are still quite similar to SQL, meaning that developers still need to learn new programming concepts.

**Improving performance of database programs by code transformation.** There is also work in terms of improving application performance by transforming loops and reordering statements to expose opportunities for query batching [9, 26], and pushing computations into the database to improve application performance [8]. Our work is orthogonal to this line of research. After converting portions of the source code into SQL queries, such code transformation tools can still be applied to gain additional performance improvement.

## 10. Conclusions

In this paper, we presented the QBS algorithm for inferring relational specifications from imperative code that retrieves data using ORM libraries. Our algorithm works by automatically inferring loop invariants and postconditions associated with the source program, and converting the validated postcondition into SQL queries. Our approach is both sound and precise in preserving the ordering of records. In developing the algorithm, we designed a new theory of ordered relations that allows efficient encoding of relational operations into a predicate language. We implemented a prototype of the algorithm, and demonstrated the applicability and limits using a variety of benchmarks from both real-world applications and manually-constructed examples.

## A. Benchmark Details

In this section we describe the benchmarks from real-world examples that used in Sec. 8.1. We group the benchmarks according to the type of relational operation each benchmark performs below.

**itracker benchmarks**

| Java Class Name | Line | Operation |
|---|---|---|
| EditProjectFormActionUtil | 219 | F |
| IssueServiceImpl | 1436 | D |
| IssueServiceImpl | 1456 | L* |
| IssueServiceImpl | 1567 | C* |
| IssueServiceImpl | 1583 | M |
| IssueServiceImpl | 1592 | M |
| IssueServiceImpl | 1601 | M |
| IssueServiceImpl | 1422 | D |
| ListProjectsAction | 70 | N* |
| MoveIssueFormAction | 144 | K* |
| NotificationServiceImpl | 561 | O |
| NotificationServiceImpl | 842 | A |
| NotificationServiceImpl | 941 | H |
| NotificationServiceImpl | 244 | O |
| UserServiceImpl | 155 | M |
| UserServiceImpl | 412 | A |

**wilos benchmarks**

| Java Class Name | Line | Operation |
|---|---|---|
| ActivityService | 401 | A |
| ActivityService | 328 | A |
| AffectedtoDao | 13 | B |
| ConcreteActivityDao | 139 | C* |
| ConcreteActivityService | 136 | D |
| ConcreteRoleAffectationService | 55 | E |
| ConcreteRoleDescriptorService | 181 | F |
| ConcreteWorkBreakdownElementService | 55 | G* |
| ConcreteWorkProductDescriptorService | 245 | F |
| GuidanceService | 140 | A |
| GuidanceService | 154 | A |
| IterationService | 102 | A |
| LoginService | 103 | H |
| LoginService | 83 | H |
| ParticipantBean | 1079 | B |
| ParticipantBean | 681 | H |
| ParticipantService | 146 | E |
| ParticipantService | 119 | E |
| ParticipantService | 266 | F |
| PhaseService | 98 | A |
| ProcessBean | 248 | H |
| ProcessManagerBean | 243 | B |
| ProjectService | 266 | K* |
| ProjectService | 297 | A |
| ProjectService | 338 | G* |
| ProjectService | 394 | A |
| ProjectService | 410 | A |
| ProjectService | 248 | H |
| RoleDao | 15 | I* |
| RoleService | 15 | E |
| WilosUserBean | 717 | B |
| WorkProductsExpTableBean | 990 | B |
| WorkProductExpTableBean | 974 | J |

where:

A: selection of records

B: return literal based on result size

C: retrieve the max / min record by first sorting and then returning the last element*

D: projection / selection of records and return results as a set

E: nested-loop join followed by projection

F: join using `contains`

G: type-based record selection*

H: check for record existence in list

I: record selection and only return the one of the records if multiple ones fulfill the selection criteria*

J: record selection followed by count

K: sort records using a custom comparator*

L: projection of records and return results as an array*

M: return result set size

N: record selection and in-place removal of records*

O: retrieve the max / min record

\* indicates those that are not currently handled by the QBS algorithm